\RequirePackage{fix-cm}
\documentclass{svjour3}          
\smartqed  
\usepackage{graphicx}
\setkeys{Gin}{clip=true,draft=false}
\graphicspath{{./pdf/}}
\DeclareGraphicsExtensions{.pdf}
\usepackage{mathptmx}      
\usepackage{cite}
\usepackage[cmex10]{amsmath}
\usepackage{amsfonts}
\usepackage[cspex,bbgreekl]{mathbbol}
\makeatletter
\def\vec@style{\relax} 
\def\vec#1{\relax\ifmmode\mathchoice
{\mbox{\boldmath$\vec@style\displaystyle#1$}}
{\mbox{\boldmath$\vec@style\textstyle#1$}}
{\mbox{\boldmath$\vec@style\scriptstyle#1$}}
{\mbox{\boldmath$\vec@style\scriptscriptstyle#1$}}\else
\hbox{\boldmath$\vec@style\textstyle#1$}\fi}

\def\mat@style{\sf} 

\def\mat#1{\relax\ifmmode\mathchoice
{\mbox{\boldmath$\mat@style\displaystyle#1$}}
{\mbox{\boldmath$\mat@style\textstyle#1$}}
{\mbox{\boldmath$\mat@style\scriptstyle#1$}}
{\mbox{\boldmath$\mat@style\scriptscriptstyle#1$}}\else
\hbox{\boldmath$\mat@style\textstyle#1$}\fi}
\def\IEEEproofnamesk{Sketch of proof}
\def\IEEEproofsk{\@ifnextchar[{\@IEEEproofsk}{\@IEEEproofsk[\IEEEproofnamesk]}}
\def\@IEEEproofsk[#1]{\par\noindent\hspace{2em}{\itshape #1: }}

\makeatother
\newcommand{\R}{{\ifmmode\mathbb{R}\else$\mathbb{R}$\fi}}
\newcommand{\C}{{\ifmmode\mathbb{C}\else$\mathbb{C}$\fi}}\newcommand{\D}{\text{d}}

\newcommand{\I}{\text{i}}

\newcommand{\gei}{{\I}}

\newcommand{\fracd}[2]{\frac{\displaystyle #1}{\displaystyle #2}}
\newcommand{\partialder}[2]{\frac{\displaystyle\partial #1}{\displaystyle\partial #2}}
\newcommand{\totder}[2]{\frac{\displaystyle\D #1}{\displaystyle\D #2}}
\newcommand{\partialdersec}[2]{\frac{\displaystyle\partial^2 #1}{\displaystyle\partial #2^2}}

\newcommand{\up}[1]{{\text{#1}}}

\newcommand{\Xvec}{\mathbf{X}}

\newcommand{\fvec}{\mathbf{f}}
\newcommand{\avec}{\mathbf{a}}

\newcommand{\Ymat}{\mathbf{Y}}
\newcommand{\Yvec}{\mathbf{Y}}

\newcommand{\Vmat}{\mathbf{V}}

\newcommand{\Bmat}{\mathbf{B}}

\newcommand{\Rmat}{\mathbf{R}}
\newcommand{\Rvec}{\mathbf{R}}
\newcommand{\Dmat}{\mathbf{D}}
\newcommand{\Umat}{\mathbf{U}}

\newcommand{\Amat}{\mathbf{A}}

\newcommand{\Zmat}{\mathbf{Z}}

\newcommand{\xvec}{\mathbf{x}}
\newcommand{\vvec}{\mathbf{v}}
\newcommand{\yvec}{\mathbf{y}}
\newcommand{\zvec}{\mathbf{z}}
\newcommand{\uvec}{\mathbf{u}}
\newcommand{\hvec}{\mathbf{h}}

\newcommand{\Wvec}{\mathbf{W}}
\newcommand{\Lvec}{\mathbf{L}}

\newcommand{\xivec}{\boldsymbol{\xi}}
\newcommand{\zerovec}{\mathbf{0}}

\newcommand{\Smat}{\mathbf{S}}

\newcommand{\Fmat}{\mathbf{F}}

\newcommand{\DIAG}{\operatorname{\text{diag}}}
\newcommand{\diag}[1]{{\DIAG\left\{#1 \right\}}}

\newcommand{\norm}[1]{{\left\| #1 \right\|}}
\newcommand{\ave}[1]{{\left\langle #1 \right\rangle}}

\journalname{}
\begin{document}

\title{Noise in oscillators: a review of state space decomposition approaches
}


\author{F.L. Traversa         \and
        M. Bonnin \and F. Corinto \and F. Bonani
}


\institute{Politecnico di Torino \\ Dipartimento di Elettronica e Telecomunicazioni\at
              Corso Duca degli Abruzzi 24, 10129 Torino, Italy \\
              \email{fabio.traversa@polito.it}           
}

\date{Received: date / Accepted: date}

\maketitle

\begin{abstract}
We review the state space decomposition techniques for the assessment of the noise properties of autonomous oscillators, a topic of great practical and theoretical importance for many applications in many different fields, from electronics, to optics, to biology. After presenting a rigorous definition of phase, given in terms of the autonomous system isochrons, we provide a generalized projection technique that allows to decompose the oscillator fluctuations in terms of phase and amplitude noise, pointing out that the very definition of phase (and orbital) deviations depends of the base chosen to define the aforementioned projection. After reviewing the most advanced theories for phase noise, based on the use of the Floquet basis and of the reduction of the projected model by neglecting the orbital fluctuations, we discuss the intricacies of the phase reduction process pointing out the presence of possible variations of the noisy oscillator frequency due to amplitude-related effects.
\end{abstract}

\section{Introduction}
\label{intro}

Since the very beginning of the history of telecommunication systems, oscillators have played a major role as frequency sources whose precision deeply impacts the system performance, especially for high sensitivity receivers. On the other hand, autonomous systems (mostly ring oscillators) are fundamental to provide the time reference required by the operation of synchronous digital systems. For both applications, oscillator noise properties define the ultimate system performance, since phase noise is the major component of frequency fluctuations, as well as of the time jitter that impairs the digital circuit synchronization. As a consequence, noise in oscillators has been studied since the dawn of electronic technology, mainly with the aim of minimizing phase noise. A wide variety of modeling techniques, many mainly oriented to circuit design are available: a recent and thorough review can be found in \cite{Pankratz}. Electronic applications are not however the only application of autonomous systems: they are also the mathematical formulation for many single neuron biological systems \cite{Wedgwood,Winfree}.

In many cases, the orbit representing the deterministic solution of the unperturbed system is strongly stable, and therefore in presence of weak random perturbations, namely noise, the resulting perturbation of the limit cycle is well described by the phase concept only. On the other hand, for some autonomous systems where the periodic orbit is only weakly stable, such as for instance Chua's circuit \cite{BonaniGilli} or some biological systems (e.g., the Morris-Lecar model) \cite{Wedgwood}, the amplitude part of oscillator noise may also play an important role. The simultaneous assessment of phase and amplitude noise in terms of a rigorous oscillator noise theory requires the decomposition of the stochastic equations modeling fluctuations along the two possible components, giving rise to the state space decomposition approaches \cite{Kaertner90,noiFN,noi,BonninNOLTA}.

The state space description of the noisy oscillator is provided in Section~\ref{1sec}, while we discuss the concept of phase and of the phase and orbital noise decomposition for oscillator fluctuations in Section~\ref{PhaseSec}. The perturbative solution approaches are presented in Section~\ref{2sec}, while the state decomposition approaches are introduced in Section~\ref{3sec} in a generalized way providing a novel derivation of the phase and amplitude decomposition  together with an accurate discussion of the reduction to phase models, pointing out the existence of additive terms that may impact on the average time reference fluctuation. This matter is further discussed in Section~\ref{FPlownoisesec} in terms of the phase fluctuation Fokker-Planck equation. Section~\ref{4sec} is finally devoted to an example of application.

\section{Noisy oscillator description in state space}
\label{1sec}

Oscillators are, from the mathematical standpoint, represented by autonomous dynamical systems, i.e. systems in which no time-varying external driving signal is applied. Of course, from a circuit standpoint, the energy required to sustain the oscillation is provided by a time-invariant (DC) generator. The second requirement is that such an autonomous dynamical admits of a periodic solution: we denote the period with $T$, the corresponding frequency is $f_0=1/T$, while the angular frequency is $\omega_0=2\pi/T$.

For the sake of simplicity, we consider a noiseless oscillator described by an Ordinary Differential Equation (ODE) of the form
\begin{equation}
\totder{\xvec}{t}=\fvec(\xvec(t)),
\label{nosc}
\end{equation}
where $\xvec(t)\in\mathbb{R}^n$ is the collection of the variables describing the system state, and $\fvec\in\mathbb{R}^n$ is a nonlinear function. The condition on the periodic solution means that \eqref{nosc} admits of a $T$ periodic orbit (a limit cycle, see e.g. \cite{Perko} for a precise definition), i.e. a function $\xvec_\up{S}(t)$ exists that satisfies \eqref{nosc} with the property $\xvec_\up{S}(t+T)=\xvec_\up{S}(t)$ $\forall t$.

Notice that this simplified choice is not representing all the circuit oscillators, since in the general case the mathematical representation of an autonomous circuit equations is an index-1 differential-algebraic equation (DAE) \cite{Sangiovanni} and not a simple ODE. The extension of our treatment to the DAE case requires a significant increase in the formal complexity, therefore we stick to simpler ODE case.

Fluctuations are induced in the real-world oscillator by the noise sources that are unavoidably present inside the autonomous system components, such as the thermal, shot and low-frequency (e.g., flicker) noise taking place in the semiconductor devices exploited for the realization of circuit oscillators \cite{libro}. Noise is represented by a collection $\xivec(t)\in\mathbb{R}^m$ of stochastic processes, that perturb \eqref{nosc} into
\begin{equation}
\totder{\Xvec}{t}=\fvec(\Xvec(t),\xivec(t)).
\label{osc1}
\end{equation}
The standard assumption is that $\xivec(t)$ are uncorrelated Gaussian random processes \cite{Demir1}, that for the sake of simplicity we assume here as white gaussian noise.

In most of the cases, \eqref{osc1} is further approximated by linearization around the noiseless solution
\begin{equation}
\totder{\Xvec}{t}\approx \fvec(\Xvec(t),\zerovec)+\left.\partialder{\fvec}{\xivec}\right|_{\xivec=\zerovec}\xivec(t)
\label{osc2}
\end{equation}
 where the Jacobian of $\fvec$ is calculated in the solution $\Xvec(t)$. Equation \eqref{osc2}, a stochastic ordinary differential equation (S-ODE) \cite{Oksendal}, is often casted in the form of a Langevin equation
\begin{equation}
\totder{\Xvec}{t}= \fvec(\Xvec(t))+\epsilon\Bmat(\Xvec(t))\xivec(t)
\label{osc3}
\end{equation}
where $\epsilon$ is a not necessarily small parameter that measures the strength of the noise sources, and $\Bmat(\Xvec(t))$ is modulating matrix that takes into account possible cyclostationary (modulated) fluctuations \cite{Demir1}. In the mathematical literature, the S-ODE \eqref{osc3} is often written as
\begin{equation}
\D\Xvec(t) = \fvec(\Xvec(t))~\D t+\epsilon\Bmat(\Xvec(t))\circ\D\Wvec(t)
\label{osc4}
\end{equation}
where $\Wvec$ is an $m$ dimensional Brownian motion (with uncorrelated components). The S-ODE \eqref{osc4} has different solutions according to whether it is interpreted in the Stra\-to\-no\-vich or It\^{o} sense \cite{Oksendal}: this matter is discussed in \cite{BonninTCAS,BonninNOLTA} in detail. In particular, the Stra\-to\-no\-vich interpretation allows for simpler theoretical derivations, since the customary calculus rules apply, and is widely considered the interpretation closer to a physically sound interpretation. On the other hand, It\^{o} interpretation is necessary for numerical solutions because of the non-anticipating nature of It\^{o} stochastic integral. Conforming to the standard notation, we use the symbol $\Bmat(\Xvec(t))\circ\D\Wvec(t)$ to denote Stra\-to\-no\-vich interpretation (as we have done in \eqref{osc4}) while we reserve notation $\Bmat(\Xvec(t))\D\Wvec(t)$ for It\^o interpretation (compare with \eqref{osc4Strato} below).

Notice that in case of Stra\-to\-no\-vich interpretation, a transformation exists \cite[Eq. (4.3.43)]{Gardiner} that transforms \eqref{osc4} into an equivalent It\^{o} equation, i.e. a It\^{o} S-ODE having the same solution. The equivalent It\^{o} equation has the same functional form of \eqref{osc4}:
\begin{equation}
\D\Xvec(t) = \hat\fvec(\Xvec(t))~\D t+\epsilon\Bmat(\Xvec(t))~\D\Wvec(t)
\label{osc4Strato}
\end{equation}
where  the first term on the right hand side (r.h.s.) is given by ($f_i$ is the $i$-th component of $\fvec$, $B_{ij}$ the $(i,j)$ element of $\Bmat$)
\begin{equation}
\hat f_i(\Xvec(t))= f_i(\Xvec(t))+\fracd{\epsilon^2}{2} \sum_k \partialder{\Bmat}{R_k} \left(\Bmat^\up{T}\right)_k
\end{equation}
where $\left(\Bmat^\up{T}\right)_k$ denotes the $k$-th column of matrix $\Bmat^\up{T}$ (i.e., the $k$-th row of $\Bmat$). The deterministic correction term on the r.h.s. of the previous equation is often called the \textit{noise induced drift term}. Since the solutions of \eqref{osc4} and \eqref{osc4Strato} are the same, also their statistical properties coincide, although the same might not hold separately for the phase and amplitude fluctuations defined according to the projection procedure discussed in Section~\ref{phamprojsec}.

Notice also that the drift term is proportional to $\epsilon^2$, thus in the low noise limit and assuming that the derivatives of $\Bmat$ with respect to the state variables are small, the two interpretations are equivalent \cite{Kaertner90}.

\section{Phase definition, phase fluctuations and orbital noise}
\label{PhaseSec}

The most important parameter to be defined in the analysis of oscillator noise is the phase concept \cite{Guckenheimer75}. The reason for this is that in the vast majority of the practical electronic applications the oscillator limit cycle is a stable solution of \eqref{nosc}, meaning that perturbations of the cycle that drive the solution $\Xvec(t)$ away from the orbit fall (often rapidly) back to the cycle itself. Therefore, most of the fluctuations take place \textit{along} the limit cycle $\xvec_\up{S}(t)$, thus leading to the phase noise concept.

\begin{figure}
\centerline{
\includegraphics[width=.9\columnwidth]{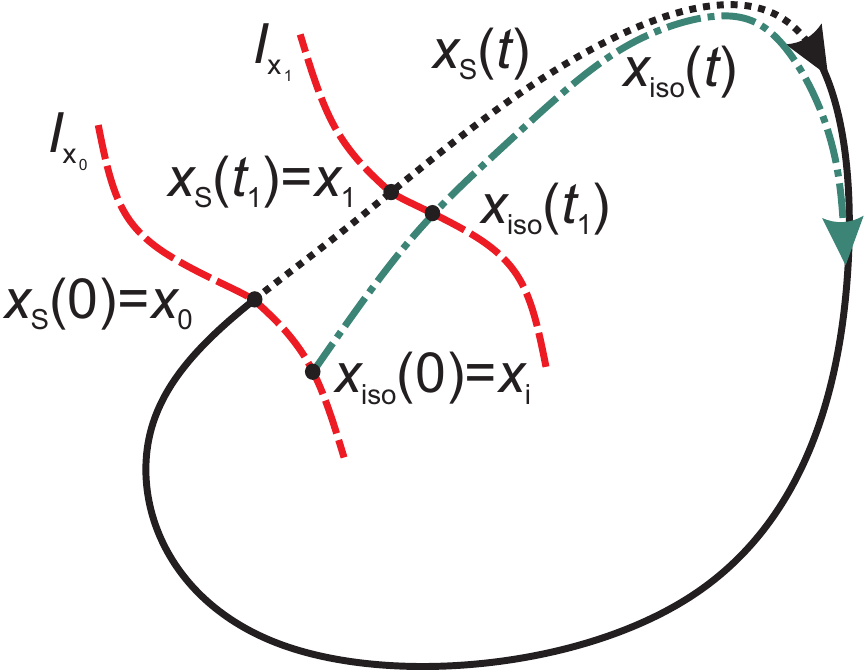}}
\caption{\label{phase_noiseless} Definition of isochron.}
\end{figure}

Let us consider first the noiseless limit cycle. Since the oscillator is an autonomous system, the orbit  is followed irrespective of the starting point $\xvec_\up{S}(0)=\xvec_0$, i.e. the time reference of a noiseless oscillator is not fixed \textit{a priori} as it happens for a forced circuit \cite{Demir1}. The phase of the point $\xvec_\up{S}(t)$ is then defined as $
\Phi_\up{d}(\xvec_\up{S}(t))=\omega_0t$. In other words, the phase defines a local coordinate on the orbit.

In presence of perturbations, e.g. when fluctuations take place, the solution $\Xvec(t)$ is no longer guaranteed to always remain on the limit cycle, thus a proper definition of phase (of course consistent with the definition given above for the unperturbed stable solution) requires to make use of the concept of \textit{isochron} \cite{Djurhuus,Suvak,BonninTCAS,BonninNOLTA}. We avoid here the details of the mathematical definition, and try to provide an operative definition and state the properties of the isochrons: let us consider a  solution of \eqref{nosc} (or, equivalently, a solution of \eqref{osc4} in the absence of noise) starting from the initial condition $\xvec_\up{S}(0)=\xvec_0$. The isochron associated to $\xvec_0$, that we denote as $I_{\xvec_0}$, is the collection of points $\xvec_\up{i}$ (in the domain of attraction of the limit cycle) such that the solution $\xvec_\up{iso}(t)$ of \eqref{osc4} starting from the initial condition $\xvec_\up{iso}(0)=\xvec_\up{i}$ asymptotically converges towards $\xvec_\up{S}(t)$, i.e.
\begin{equation}
\lim_{t\rightarrow+\infty} \norm{\xvec_\up{iso}(t)-\xvec_\up{S}(t)}=0
\end{equation}
where $\norm{\cdot}$ is a properly defined norm. For a geometrical interpretation in 2D, see Fig.~\ref{phase_noiseless}.

For each oscillator of size $n$, the isochrons form a hyper-surface in $\mathbb{R}^n$ of size $n-1$, and they are shown \cite{Farkas,Suvak} to be endowed of the following property: trajectories starting from the same isochron cross other isochrons at the same time instant (Fig.~\ref{phase_noiseless}). Accordingly, we define the phase of a point $\xvec_\up{i}\in I_{\xvec_0}$ in the basin of attraction of the stable orbit by assigning the same value to all the points lying on $I_{\xvec_0}$:
\begin{equation}
\Phi_\up{i}(\xvec_\up{iso}(t))=\Phi_\up{i}(\xvec_i)+\omega_0t.
\label{moja}
\end{equation}
Combining \eqref{osc1} and \eqref{moja}, we  find
\begin{equation}
\totder{\Phi_\up{i}(\xvec)}{t}=\nabla_\xvec\Phi_\up{i}(\xvec)\totder{\xvec}{t}=\nabla_\xvec\Phi_\up{i}(\xvec)\fvec(\xvec)=\omega_0.
\label{mbili}
\end{equation}

In presence of noise,  the phase $\Phi_\up{i}(\Xvec)$ becomes a stochastic variable, leading to the oscillator \textit{phase noise}. Since $\Phi_\up{i}(\Xvec)$ is not a complete representation of the oscillator state but rather a 1D variable, phase noise is not a complete representation of the noisy oscillator. However, since practical oscillators are almost invariable strongly stable systems, phase noise contains the ``majority'' of the fluctuations and therefore is often considered ``the'' expression of oscillator noise. Nevertheless, as we will see later on, in many cases the other noise components may provide a significant, albeit rarely dominant, contribution.

The other $n-1$ fluctuating variables required to complete the characterization, that we shall collectively denote as the ($n-1$ components) vector $\Rvec$,  constitute the oscillator \textit{amplitude} (or \textit{orbital}) \textit{noise}. In mathematical terms, we seek an invertible variable transformation $\hvec(\Phi,\Rvec)$ such that $\Xvec=\hvec(\Phi,\Rvec)$: different choices for such a transformation lead to different decompositions into phase and orbital noise. A discussion on this will be provided in Section~\ref{phasereducedsec}.

To allow for an easier comparison with the available literature, we remark that often the phase of the noisy solution is expressed as a function of a zero-average time reference fluctuation $\alpha(t)$ \cite{Demir1} (notice that in \cite{Kaertner90} symbol $\theta(t)$ is used, however is this work we stick to the notation proposed in \cite{Demir1}), linked to $\Phi(t)$ by
\begin{equation}
\Phi(t)=\omega_0(t+\alpha(t)).
\label{alfadef}
\end{equation}

\section{S-ODE solution: perturbative approaches}
\label{2sec}

Several techniques have been used in the long history of oscillator noise analysis for the solution of \eqref{osc3} (or, equivalently, \eqref{osc4}). In many cases, the effect of the noise sources is expressed as a perturbation of the noiseless solution
\begin{equation}
\Xvec(t)=\xvec_\up{S}(t)+\xvec_\up{n}(t),
\label{tatu}
\end{equation}
where $\xvec_\up{n}(t)$ is a zero-average stochastic process. The two-time correlation matrix of $\xvec_\up{n}(t)$ ($\ave{\cdot}$ denotes the statistical average)
\begin{equation}
\Rmat_{\xvec_\up{n},\xvec_\up{n}}(t_1,t_2)=\ave{\xvec_\up{n}(t_1)\xvec_\up{n}(t_2)}
\end{equation}
defines the oscillator state noise.

The simplest solution approach for \eqref{osc4} consists in assuming that the noise sources linearly perturb the noiseless solution, i.e. $\xvec_\up{n}(t)$ is a small perturbation. In this case, \eqref{osc4} is linearized either around a properly chosen time-independent (namely, DC) value, or more generally around the noiseless solution. In the former case, the linearized relationship is time invariant and therefore the corresponding modeling approach is termed LTI (linear time invariant) \cite{Hajimiribook}. On the other hand, linearizing around the limit cycle leads to a periodically time-varying linear system that is denoted as LTV (linear time varying) approach \cite{Hajimiribook}.

While the LTI is often poorly accurate, the LTV technique is widely popular because it combines a relative simplicity of the approach and a good accuracy, at least as far as spectrum frequencies not too close to the fundamental $f_0$ and to its integer multiples are considered. In fact, all these approaches are plagued by the same fundamental flaw: they provide a diverging (infinite) spectrum for $f\rightarrow kf_0$ ($k$ integer). On the other hand, because of the linearization, in most of the circuits an analytical link between circuit parameters and phase noise can be derived, thus making these approaches strongly design oriented.

\section{S-ODE solution: state space decomposition approaches}
\label{3sec}

Despite their simplicity, the accuracy of the linearized perturbative approaches discussed above has been often questioned, in particular for frequencies close to the oscillation fundamental \cite{Demir1}. This led to the development of more mathematically founded, and therefore less design-oriented, theories. In many cases, a S-ODE with various degrees of approximations is derived, and then studied ultimately by characterizing the statistical properties of the phase fluctuations exploiting a deterministic partial differential equation (PDE) whose unknown is the probability density function of the phase fluctuation itself: such a PDE is the Fokker-Planck equation associated to the S-ODE \cite{Gardiner,Risken}. Several examples of these approaches can be found in \cite{Pankratz,Lax,Kaertner89,Kaertner90,Demir1,Suvak} and in the references therein.

However, phase noise is only one of the components of oscillator noise: a complete characterization calls for the derivation of S-ODEs describing both the phase and orbital noise components. This procedure is called \textit{state decomposition}, whose starting point is the definition of the phase component of the noisy state $\Xvec(t)$.

An important step in the derivation of a state space decomposition is provided by the seminal papers by F. Kaertner \cite{Kaertner89,Kaertner90}, that laid the basis for the following developments leading to the Floquet-based theories in \cite{Demir1,Demir3,noi} (Floquet theory is briefly introduced in Appendix~\ref{AppB}). The basic idea is to recognize the phase noise contribution as the projection of the solution of the S-ODE \eqref{osc3} \textit{along} the noiseless solution $\xvec_\up{S}(t)$, while the remaining portion of the solution space defines the amplitude (orbital) fluctuation.

Let us consider first the unit vector tangent to the limit cycle:
\begin{equation}
\uvec_1(t)=\fracd{1}{\norm{\D\xvec_\up{S}/\D t}}\totder{\xvec_\up{S}}{t},
\label{normu1}
\end{equation}
that coincides with the (normalized) direct Floquet eigenvector associated to the unitary Floquet multiplier \cite{Demir1}. We also consider for the moment $n-1$ other vectors $\uvec_2(t),\dots,\uvec_n(t)$ having the only property of being linearly independent among them and with $\uvec_1(t)$. Therefore, $\{\uvec_1(t),\dots,\uvec_n(t)\}$ spans (i.e., is a basis) for $\mathbb{R}^n$ and therefore the square matrix $\Umat(t)$ whose columns are vectors $\uvec_1(t),\dots,\uvec_n(t)$ is invertible. The inverse is denoted as $\Vmat(t)=\Umat^{-1}(t)$, allowing to define a set of $n$ vectors $\{\vvec_1(t),\dots,\vvec_n(t)\}$ ($^\up{T}$ denotes the transpose)
\begin{equation}
\Vmat(t)=\Umat^{-1}(t)=\begin{bmatrix}
\vvec_1^\up{T}(t) \\ \vdots \\ \vvec_n^\up{T}(t)
\end{bmatrix}
\end{equation}
that by definition are bi-orthogonal with $\uvec_i(t)$, i.e.
\begin{equation}
\vvec_i^\up{T}(t)\uvec_j(t)=\delta_{i,j}\qquad \forall i,j=1,\dots,n
\label{biob}
\end{equation}
where $\delta_{i,j}$ is Kronecker symbol. Clearly, also $\{\vvec_1(t),\dots,\vvec_n(t)\}$ is a basis for $\mathbb{R}^n$.

We will exploit these general bi-orthogonal bases to derive in general terms an accurate state state decomposition that correctly includes all the terms of the same order in the noise strength coefficient $\epsilon$, thus improving the decomposition introduced in \cite{Kaertner90} and then solved exactly for the phase part in \cite{Demir1} and for the amplitude part in \cite{noi}. Notice that in the latter contributions the bi-orthogonal bases were chosen starting from the direct and adjoint Floquet eigenvectors discussed in Appendix~\ref{AppB}. 

\section{The novel S-ODE  state space decomposition}
\label{phamprojsec}

Starting from the bi-orthogonal bases, a projection procedure allows for the determination of the S-ODEs that define the phase and amplitude fluctuations.

Let us introduce first two rectangular matrices made each of $n-1$ vectors of the bases $\{\uvec_i(t)\}$ and $\{\vvec_i(t)\}$:
\begin{equation}
\Ymat(\Phi)=[\uvec_2(\Phi),\dots,\uvec_n(\Phi)],\quad \Zmat(\Phi)=[\vvec_2(\Phi),\dots,\vvec_n(\Phi)].
\label{mama}
\end{equation}
Notice that at this level, the bases do not need to be made of the Floquet eigenvectors. The only requirement is that they both span the subspace on $\mathbb{R}^{n-1}$ that is linearly independent of $\uvec_1(t)$.

In order to take advantage of the easier calculus rules, we  use the Stra\-to\-no\-vich interpretation of \eqref{osc4}, and decompose the state stochastic variable in the same way as in \cite{Kaertner89,Kaertner90,Demir1}
\begin{equation}
\Xvec(t)=\hvec(\Phi,\Rvec)=\xvec_\up{S}(\Phi)+\Ymat(\Phi)\Rvec(t).
\label{punda}
\end{equation}
Notice that to guarantee that if $\Rvec(t)$ is small, this condition holds irrespective of the normalization chosen for $\Ymat$, the column vectors $\uvec_k(t)$ $k=2,\dots,n$ should be chosen normalized to 1.

To derive the stochastic equations for the phase and amplitude, we compute the derivative of $\Xvec$, thus finding (the explicit dependence on $\Phi$ and $t$ is omitted for simplicity)
\begin{align}
\D\Xvec & =  \left( \partialder{\xvec_\up{S}}{\Phi} + \partialder{\Ymat}{\Phi} \Rvec \right) \D \Phi + \Ymat \D \Rvec.\label{ItodX}
\end{align}
Projecting the \eqref{ItodX} onto the linear space  spanned by $\vvec_1(\Phi)$ we obtain the phase equation, while the projection onto the space spanned by the columns of $\Zmat(\Phi)$ defines the amplitude dynamics. 
The resulting phase and amplitude S-ODEs are:
\begin{subequations}
\label{projosc}
\begin{align}
\D\Phi &=\omega_0\left[ 1+a_\Phi(\Phi,\Rvec)\right]~\D t
+\epsilon\omega_0\Bmat_\Phi(\Phi,\Rvec)\circ\D\Wvec(t) \label{phaseexact} \\[1ex]
\D\Rvec&=\left[ \Lvec(\Phi)\Rvec+\avec_R(\Phi,\Rvec)\right]~\D t
+\epsilon\Bmat_R(\Phi,\Rvec)\circ\D\Wvec(t). \label{Rexact}
\end{align}
\end{subequations}
where
\begin{align}
&a_\Phi(\Phi,\Rvec)=\left[ r+\vvec_1^\up{T}\partialder{\Ymat}{\Phi}\Rvec\right]^{-1} \vvec_1^\up{T} \nonumber\\
&\quad\times\left[ \fvec(\xvec_\up{S}+\Ymat\Rvec)-\fvec(\xvec_\up{S})-\partialder{\Ymat}{\Phi}\Rvec \right] \label{prima} \\[1ex]
&\Bmat_\Phi(\Phi,\Rvec)=\left[ r+\vvec_1^\up{T}\partialder{\Ymat}{\Phi}\Rvec\right]^{-1} \vvec_1^\up{T} \Bmat(\xvec_\up{S}+\Ymat\Rvec)\\[1ex]
&\Lvec(\Phi)=-\Zmat^\up{T}\partialder{\Yvec}{\Rvec}\\[1ex]
&\avec_R(\Phi,\Rvec)=-\Zmat^\up{T}\left[ \partialder{\Yvec}{\Phi}\Rvec a_\Phi-\fvec(\xvec_\up{S}+\Ymat\Rvec) \right]\\[1ex]
&\Bmat_R(\Phi,\Rvec)=\Zmat^\up{T}\Bmat(\xvec_\up{S}+\Ymat\Rvec)-\Zmat^\up{T}\partialder{\Ymat}{\Phi}\Rvec \Bmat_\Phi(\xvec_\up{S}+\Ymat\Rvec) \label{ultima}
\end{align}
and $r(\Phi)=\|\fvec(\xvec_\up{S}(\Phi))\|$.

\subsection{Phase reduced model}
\label{phasereducedsec}

Since in most of the applications phase noise is the dominant fluctuation component, we seek a reduction of the complete state space decomposition \eqref{projosc} into a phase reduced model, i.e. a single, scalar S-ODE for the phase variable. The simplest reduction approach amounts to assume that amplitude fluctuations remain confined in a neighborhood of $\Rvec=\zerovec$ consistently with the asymptotic stability of the limit cycle.  As discussed in \cite{Yoshimura,Teramae,Goldobin,BonninTCAS,BonninNOLTA}, a direct substitution of $\Rvec=\zerovec$ into the Stra\-to\-no\-vich equation system \eqref{projosc} may lead to an oversimplified phase equation, thus implying a significant loss of information.

Such a loss of information is due to the fact that in the Stra\-to\-no\-vich interpretation stochastic variables and noise increments are actually correlated \cite{Gardiner}. Simply setting $\Rvec=\zerovec$ amounts to treat the amplitude fluctuation as a parameter, thus loosing its stochastic nature together with the aforementioned correlation. In order to avoid such issue, the correlation should be removed first. This can be achieved transforming the Stra\-to\-no\-vich system \eqref{projosc} into the equivalent It\^o system. In fact, as a consequence of the non anticipating nature of It\^o stochastic integral, in It\^o equations stochastic variables and noise increments are uncorrelated.

Transforming \eqref{projosc} into the equivalent It\^o equation yields
\begin{subequations}
\label{projoscIto}
\begin{align}
\D\Phi &=\omega_0\left\{  1+a_\Phi(\Phi,\Rvec)+\fracd{\epsilon^2}{2}\left[ \omega_0\partialder{\Bmat_\Phi}{\Phi}\Bmat_\Phi^\up{T} \right.\right.\nonumber\\[1ex]
&\qquad\left.\left.+ \sum_k \partialder{\Bmat_\Phi}{R_k}\left( \Bmat_R^\up{T}\right)_k\right]\right\}~\D t + \epsilon\omega_0\Bmat_\Phi~\D\Wvec(t) \label{primaIto}\\[1ex]
\D\Rvec&=\left\{  \Lvec(\Phi)\Rvec+\avec_R(\Phi,\Rvec)+\fracd{\epsilon^2}{2}\left[ \omega_0\partialder{\Bmat_R}{\Phi}\Bmat_\Phi^\up{T} \right.\right.\nonumber\\[1ex]
&\qquad\left.\left.+\sum_k \partialder{\Bmat_R}{R_k}\left( \Bmat_R^\up{T}\right)_k\right]\right\}~\D t + \epsilon\Bmat_R~\D\Wvec(t), \label{secondaIto}
\end{align}
\end{subequations}
where the $(\Phi,\Rvec)$ dependence has been dropped from $\Bmat_\Phi$ and $\Bmat_R$, and $\left( \Bmat_R^\up{T}\right)_k$ denotes the $k$-th column of $\Bmat_R^\up{T}$.

We can now approximate \eqref{primaIto} by setting $\Rvec=\zerovec$, and simply drop \eqref{secondaIto}. Since $a_\Phi(\Phi,\zerovec)=0$, we find
\begin{align}
\D\Phi &=\omega_0\left\{ 1+\fracd{\epsilon^2}{2}
\left[ \omega_0\partialder{\Bmat_\Phi}{\Phi}\Bmat_\Phi^\up{T} + \sum_k \partialder{\Bmat_\Phi}{R_k}\left( \Bmat_R^\up{T}\right)_k\right]\right\}~\D t\nonumber\\[1ex]
&\quad + \epsilon\omega_0\Bmat_\Phi~\D\Wvec(t).
\label{phaseIto}
\end{align}
where $\Phi$ is now the only stochastic variable. Turning back to the Stra\-to\-no\-vich equation, \eqref{phaseIto} becomes
\begin{equation}
\D\Phi =\omega_0\left\{ 1+\fracd{\epsilon^2}{2}
 \sum_k \partialder{\Bmat_\Phi}{R_k}\left( \Bmat_R^\up{T}\right)_k\right\}~\D t
 + \epsilon\omega_0\Bmat_\Phi\circ\D\Wvec(t).
\label{phaseStrato}
\end{equation}

To compare with the literature, we use as a stochastic variable the time fluctuation $\alpha(t)$ defined in \eqref{alfadef}. The phase equation becomes
\begin{equation}
\D\alpha =\fracd{\epsilon^2}{2}
 \sum_k \partialder{\Bmat_\Phi}{R_k}\left( \Bmat_R^\up{T}\right)_k~\D t
 + \epsilon\Bmat_\Phi\circ\D\Wvec(t)
\label{timeStrato}
\end{equation}
where $\Bmat_\Phi$ and $\Bmat_R$ are now calculated in $t+\alpha(t)$.

Notice that in the vanishingly small noise limit ($\epsilon\rightarrow 0$),  the term proportional to $\epsilon^2$ can be neglected thus obtaining
\begin{equation}
\D\alpha = \epsilon\Bmat_\Phi\circ\D\Wvec(t).
\label{DemirStrato}
\end{equation}
Equation \eqref{DemirStrato} coincides with the phase equation derived in \cite{Demir1} (apart from the $\epsilon$ factor, that in \cite{Demir1} is included in $\Bmat$), since for $\Rvec=\zerovec$
\begin{equation}
\Bmat_\Phi(t+\alpha(t),\zerovec)=r^{-1}(t+\alpha(t)) \vvec_1^\up{T}(t+\alpha(t)) \Bmat(\xvec_\up{S}(t+\alpha(t))),
\end{equation}
and the normalization constant $r^{-1}(t)$ is absorbed by the Floquet eigenvector $\vvec_1(t)$. In fact
\begin{equation}
r(t)=\norm{\fvec(\xvec_\up{S}(t))}=\norm{\totder{\xvec_\up{S}}{t}}
\end{equation}
is the quantity we have used in \eqref{normu1} to normalize the direct (and, as a consequence, adjoint) eigenvector associated to the orbit tangent, while in \cite{Demir1} $\uvec_1(t)=\D\xvec_\up{S}/\D t$.

Also F. Kaertner in \cite{Kaertner90} derived the same equation \eqref{DemirStrato} for the phase fluctuations, however before solving the S-ODE a further approximation was made. Expressing \eqref{DemirStrato} in integral form fixing $\alpha(0)=0$, the author assumed
\begin{align}
\alpha(t) &= \epsilon \int_0^{t+\alpha(t)} r^{-1} (\eta) \vvec_1^\up{T}(\eta) \Bmat(\xvec_\up{S}(\eta))\Wvec(\eta)\D\eta \nonumber\\[1ex]
&\approx \epsilon \int_0^{t} r^{-1} (\eta) \vvec_1^\up{T}(\eta) \Bmat(\xvec_\up{S}(\eta))\Wvec(\eta)\D\eta.
\label{kaertnerint}
\end{align}
In other words, the theory in \cite{Kaertner90} corresponds to the phase S-ODE
\begin{equation}
\D \alpha = \epsilon \, r^{-1} (t) \vvec_1^\up{T}(t) \Bmat(\xvec_\up{S}(t)) \D\Wvec(t),
\label{kaertner}
\end{equation}
i.e. a linear equation with respect to the unknown $\alpha(t)$.

Notice that also the popular Impulse Sensitivity Function (ISF) technique for phase noise estimation \cite{HajimiriJSSC,Hajimiribook} can be traced back to the Kaertner simplified phase equation \eqref{kaertner}, at least as far as the ``numerical ISF'' \cite[Appendix A]{HajimiriJSSC} is considered \cite{Srivastava}.

We conclude this section with an important remark. The derivation presented here appears to require uniquely that the chosen bases are satisfying the bi-orthogonality condition \eqref{biob}. Of course, different base choices lead to different definitions for the state space decomposition \eqref{punda}, i.e. different definitions for the phase and amplitude deviations. Such an indetermination is not causing any flaw in the noise characterization as far as the full decomposition (i.e., both amplitude and phase) is used. On the other hand, problems may raise when the phase reduced model is considered. A wise choice would to choose a base that defines a phase  as close as possible to the phase function defined starting from the isochron concept in Section~\ref{PhaseSec}.  In this respect, Floquet bases guarantee a first order approximation for the isochrons \cite{Kuramoto} and thus provide a viable tool for oscillator noise analysis. In terms of the phase decomposition procedure followed above, this statement can be better understood considering that for the Floquet bases, according to Appendix~\ref{AppB}, the matrices defined in \eqref{mama} satisfy
\begin{align}
\totder{\Ymat}{t}&=\Amat(t)\Ymat(t)-\Ymat(t)\Fmat(t) \\[1ex]
\totder{\Zmat}{t}&=-\Amat^\up{T}(t)\Zmat(t)+\Zmat(t)\Fmat(t)
\end{align}
where $\Fmat=\diag{\mu_i}$ for $i=2,\dots,n$ and $\Amat(t)=\nabla_\xvec\fvec|_{\xvec_\up{S}(t)}$. Therefore, direct substitution shows that the coefficient $a_1$ defined in \eqref{prima} tends to zero quadratically with $\Rvec$.

\section{Fokker-Planck equation for phase noise in the low-noise limit}
\label{FPlownoisesec}

A viable technique to solve \eqref{timeStrato} consists of transforming the S-ODE in the associated, deterministic Fokker-Planck equation \cite{Gardiner,Risken}. The latter is a partial differential equation having as unknown the time-varying probability density function $p(\alpha,t)$ associated to the stochastic process $\alpha(t)$. We find:
\begin{align}
&\partialder{p(\alpha,t)}{t}\nonumber\\[1ex]
&\quad=-\epsilon^2\partialder{\phantom{\alpha}}{\alpha}\left\{ \left[ \partialder{\Bmat_\Phi}{\alpha} \Bmat_\Phi^\up{T} + \sum_k \partialder{\Bmat_\Phi}{R_k} \left(\Bmat_R^\up{T}\right)_k \right] p(\alpha,t) \right\} \nonumber\\[1ex]
&\qquad+\fracd{\epsilon^2}{2} \partialdersec{\phantom{\alpha}}{\alpha}\left[ \Bmat_\Phi\Bmat_\Phi^\up{T}p(\alpha,t) \right] \label{FP}
\end{align}
where all the functions are calculated in $t+\alpha$. Notice that all the terms in the right hand side of \eqref{FP} are of the same order in the noise magnitude, i.e. proportional to $\epsilon^2$, thus suggesting that the additional term in \eqref{timeStrato} might play a significant role on the stochastic features of process $\alpha(t)$, and thus on the oscillator phase noise.

We can compare \eqref{FP} with the corresponding Fokker-Planck equation derived in \cite{Demir1}\footnote{Notice that in \cite{Demir1} the Fokker-Planck equation is derived for both the Stra\-to\-no\-vich and It\^{o} interpretations: since we make use of Stra\-to\-no\-vich calculus in \eqref{DemirStrato}, equation (21) in \cite{Demir1} should be used with $\lambda=1$.}:
\begin{align}
\partialder{p(\alpha,t)}{t}
&=-\epsilon^2\partialder{\phantom{\alpha}}{\alpha}\left\{ \left[ \partialder{\Bmat_\Phi}{\alpha} \Bmat_\Phi^\up{T}  \right] p(\alpha,t) \right\} \nonumber\\[1ex]
&\qquad+\fracd{\epsilon^2}{2} \partialdersec{\phantom{\alpha}}{\alpha}\left[ \Bmat_\Phi\Bmat_\Phi^\up{T}p(\alpha,t) \right]. \label{FPDemir}
\end{align}
Clearly, the term neglected in \cite{Demir1} is of the same order of magnitude in $\epsilon$ as the last term in \eqref{FP}. This means that discarding such a term is a priori not generally a well justified assumption, even in the low noise limit, thus suggesting that the elimination of quadratic terms directly in the S-ODE is a delicate step to be taken.

\section{Example:  a comparison of phase macromodels in Coram oscillator}
\label{4sec}

We discuss here the statistical properties of the time frame fluctuation $\alpha(t)$ that derive by the application of the phase models previously proposed in literature and compare them with the reduced model \eqref{timeStrato}. As an example, we consider the Coram oscillator \cite{Coram}. The advantages of focusing on this example are twofold. First, the results for Coram system can in many cases be obtained analytically. Second, Coram oscillator is one of the few autonomous systems for which an explicit expression for the isochron-based  phase function $\Phi_\up{i}(t)$ defined in Section~\ref{PhaseSec} can actually be calculated. Therefore, we use $\Phi_\up{i}(t)$ as the reference solution.

For the sake of simplicity, we consider only the case of additive noise, i.e. we assume that $\Bmat$ is a constant matrix. In polar coordinates, noisy Coram oscillator is defined by the parameterized S-ODE system
\begin{subequations}
\label{Coram}
\begin{align}
\D \rho & = \left(\rho - \rho^2\right) \D t + \epsilon \circ \D W_{\rho}(t)  \label{VVV}\\[1ex]
\D \theta& = a_{\theta} (1 + \rho)~\D t + \epsilon \circ \D W_{\theta}(t) \label{WWW}
\end{align}
\end{subequations}
where $a_\theta$ is a real parameter.
For the noiseless system, i.e. \eqref{Coram} with $\epsilon=0$, the limit cycle is the unit circle:
\begin{equation}
\xvec_\up{S}(t)=\begin{bmatrix}
\rho_\up{S}(t)\\ \theta_\up{S}(t)
\end{bmatrix}=\begin{bmatrix}
1\\ 2a_\theta t
\end{bmatrix},
\end{equation}
while the  isochron-based phase function is shown to be \cite{BonninTCAS2}
\begin{equation}
\Phi_\up{i} =\theta + a_{\theta} \log \rho
\end{equation}
that satisfies $\D \Phi_\up{i} = 2a_\theta\D t$, i.e. $\omega_0=2a_\theta$. Therefore the level sets of $\Phi_\up{i}$ are the isochrons of the Coram system. Since the noise term is additive, \eqref{Coram} provides the same flutuations irrespective of whether it is interpreted as a It\^o or Stra\-to\-no\-vich S-ODE. Nevertheless, care must be exerted because once an interpretaion is chosen, the corresponding set of calculus rules should be used. Choosing Stra\-to\-no\-vich interpretation, and introducing the amplitude deviation $R = \rho-1$, \eqref{Coram} yields the state decomposition formulation consistent with the isochrone phase definition
\begin{subequations}
\label{CoramPhaseStrato}
\begin{align}
\D R & = -\left(R + R^2\right) \D t + \epsilon \circ \D W_{\rho}(t)\\[1ex]
\D \Phi_\up{i} & =2a_\theta \D t + \epsilon \left( \D W_{\theta}(t) + \dfrac{a_{\theta}}{R+1} \circ \D W_{\rho} \right). \label{BBB}
\end{align}
\end{subequations}
Notice that in \eqref{CoramPhaseStrato}, due to the nonlinear change of variables, noise is now multiplicative, and therefore the choice between the two possible S-ODE interpretations is no longer immaterial. Equation \eqref{CoramPhaseStrato} can now be transformed into the equivalent It\^o S-ODE obtaining
\begin{subequations}
\label{CoramPhaseIto}
\begin{align}
\D R & = -\left(R + R^2\right) \D t + \epsilon \D W_{\rho}(t)\\[1ex]
\D \Phi_\up{i}  & = 2a_\theta\left(1 - \dfrac{\epsilon^2}{4(R+1)^2} \right) \D t + \epsilon \left( \D W_{\theta}(t) + \dfrac{a_{\theta}}{
R+1} \circ \D W_{\rho} \right). \label{AAA}
\end{align}
\end{subequations}
Setting $R=0$ in \eqref{AAA} and transforming back to a  Stra\-to\-no\-vich S-ODE we find the reduced phase model
\begin{equation}
\D \Phi_\up{i}  =2a_\theta \left(1 - \dfrac{\epsilon^2}{4} \right) \D t + \epsilon \left( \D W_{\theta}(t) + a_{\theta}\circ \D W_{\rho} \right) 
\end{equation}
while in terms of the time reference deviation $\alpha_\up{i} = \Phi_\up{i}/(2a_\theta)-t$  we get
\begin{equation}
\D \alpha_\up{i} = - \dfrac{\epsilon^2}{4} \D t + \dfrac{\epsilon}{2 a_{\theta}} \left( \D W_{\theta}(t) + a_{\theta}\circ \D W_{\rho} \right). \label{newreduced}
\end{equation}
On the other hand, if the reduction $R=0$ is performed directly on  \eqref{BBB}, the phase equation in terms of the $\alpha_\up{i}(t)$ variable reads
\begin{equation}
\D \alpha_\up{i} = \dfrac{\epsilon}{2 a_{\theta}} \left( \D W_{\theta}(t) + a_{\theta}\circ \D W_{\rho} \right) .\label{oldreduced}
\end{equation}
The accuracy of the two models is assessed by comparing the results of \eqref{newreduced} and \eqref{oldreduced}. We evaluate the expected  frequency variation $\D \ave{\alpha}/\D t$. Since  noise in both \eqref{newreduced} and \eqref{oldreduced} is additive, we can take the expectation value and use the fact that Wiener processes have zero average. Clearly, the traditional reduced model  \eqref{oldreduced} yields a zero expectation value, i.e. the noisy oscillator does not show any frequency frequency shift, whereas the new reduced model \eqref{newreduced} predicts a frequency shift given by $-\epsilon^2/4$.

A more detailed comparison is available because of the possibility to analyticaly solve the Coram system. Starting from \eqref{VVV} we can write the Fokker-Planck equation for the amplitude 
\begin{equation}
\partialder{p(\rho,t)}{t} = - \partialder{\phantom{\rho}}{\rho} \left[\left(\rho-\rho^2\right) p(\rho,t) \right] + \dfrac{\epsilon^2}{2} \partialdersec{p(\rho,t)}{\rho},
\end{equation}
whose stationary solution satisfies
\begin{equation}
\partialder{\phantom{\rho}}{\rho} \left\{ \dfrac{\epsilon^2}{2} \partialder{p_{\up{st}}(\rho)}{\rho} - \left[\left(\rho-\rho^2\right) p_{\up{st}}(\rho) \right] \right\}=0.
\end{equation}
Taking into account the physically suggested boundary conditions
\begin{align}
\lim_{\rho\rightarrow 0} p_{\up{st}}(\rho) &= \lim_{\rho\rightarrow 0} \partialder{p_{\up{st}}(\rho)}{\rho} =0 \\[1ex]
\lim_{\rho\rightarrow +\infty} p_{\up{st}}(\rho) &=  \lim_{\rho\rightarrow +\infty} \partialder{p_{\up{st}}(\rho)}{\rho} = 0
\end{align}
we obtain the solution
\begin{equation}
p_{\up{st}}(\rho) = p_0 \exp\left[ \dfrac{2}{\epsilon^2}\left(\dfrac{\rho^2}{2} - \dfrac{\rho^3}{3} \right) \right]
\end{equation}
where the normalization constant $p_0$ is obtained imposing 
\begin{equation}
\int_0^{2\pi} \D \theta \int_0^{+\infty} p_\up{st}(\rho) ~ \D \rho =1.
\end{equation} 

From the stationary distribution we can calculate the expectation value
\begin{equation}
\ave{\rho} = \int_0^{2\pi} \D \theta \int_0^{+\infty} \rho p_{\up{st}}(\rho) ~\D \rho
\end{equation} 
that in turns permits to calculate the expected normalized frequency shift, defined as $\D\ave{\theta}/\D t-\omega_0$. Exploiting  \eqref{WWW} we get
\begin{equation}
\totder{\ave{\theta}}{t}-\omega_0=a_\theta(\ave{\rho}-1). \label{ZZZ}
\end{equation}

\begin{figure}
\centerline{
\includegraphics[width=.99\columnwidth]{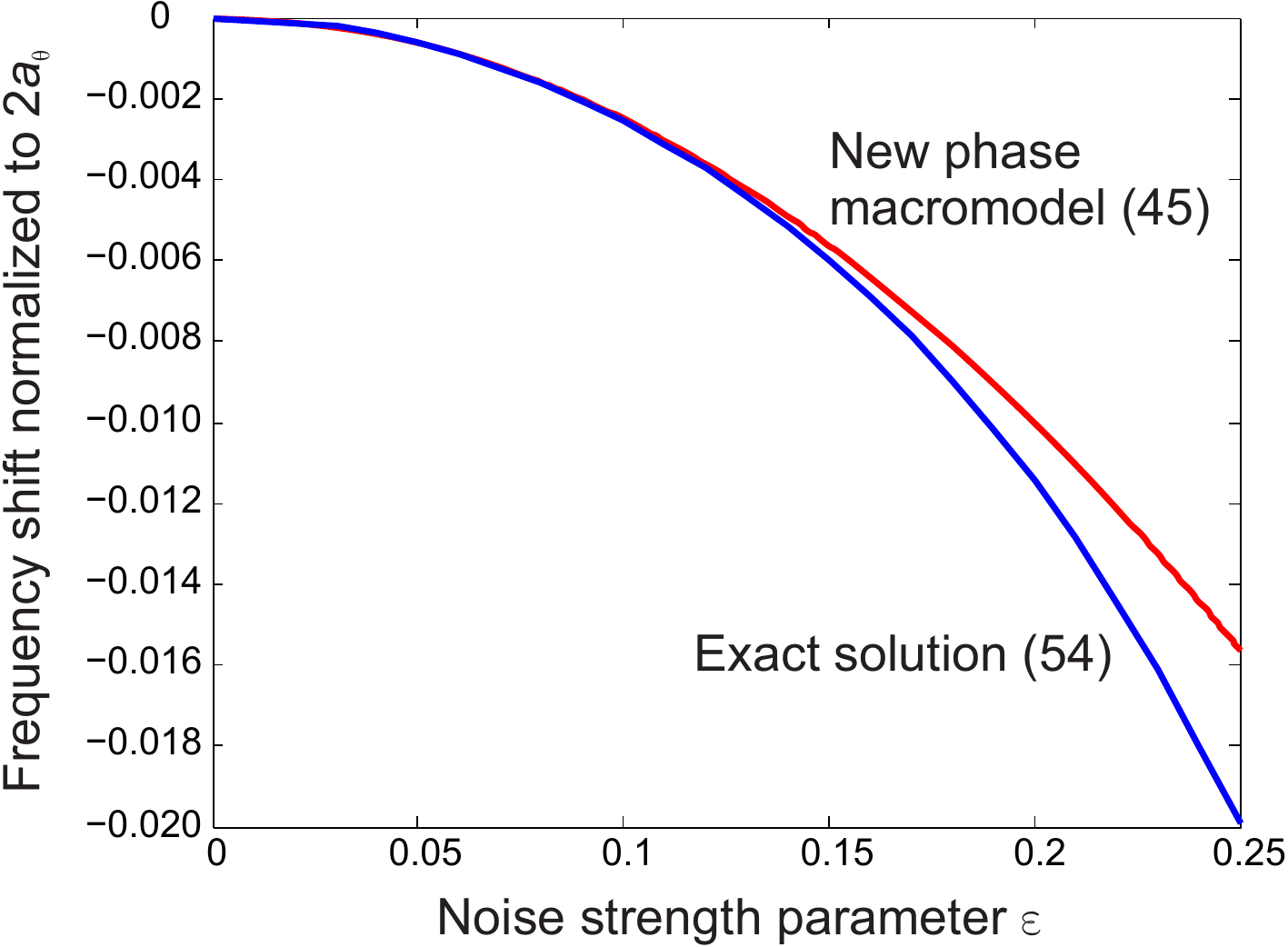}}
\caption{\label{comparison}Comparison between the exact frequency shift calculated for the Coram oscillator and the result of the novel improved phase macromodel \eqref{newreduced}.}
\end{figure}

A comparison between the exact frequency shift \eqref{ZZZ} and the prediction of our improved phase reduced model \eqref{newreduced} is shown in Fig.~\ref{comparison} as a function of the noise intensity $\epsilon$, showing that our improved phase macromodel matches the exact frequency shift for low $\epsilon$ values, and provides a good approximation for a significant range of noise strength values.

\section{Conclusion}

Starting from the rigorous definition of oscillator phase in terms of the autonomous system isochrons, we have discussed the intricacies of the derivation of phase reduced oscillator noise macromodels based on the projection of the state space equations using bi-orthogonal bases having in common the tangent to the noiseless orbit. Exploiting the Floquet bases, chosen because they represent a first order approximation of the isochrons close to the orbit (i.e., for vanishing amplitude fluctuations), we have derived the classical phase macromodels in \cite{Demir1} and \cite{Kaertner90,HajimiriJSSC}. Furthermore, we have proposed a novel phase reduction technique based on a careful discussion of the statistical properties of the solutions of S-ODE in the Stra\-to\-no\-vich and It\^o interpretations. The new macromodel leads, in general, to the prediction of a noise induced frequency variation proportional to the square of the noise strength parameter $\epsilon$. Finally, such prediction has been proven dependable in the case of the Coram oscillator, a 2D system that admits of an analytical treatment.

\begin{acknowledgements}
This work was partially supported by the Ministry of Foreign Affairs (Italy) ``Con il contributo del Ministero degli Affari Esteri, Direzione Generale per la Promozione del Sistema Paese.''
\end{acknowledgements}

\appendix

\section{Floquet theory basics}
\label{AppB}

Floquet theory forms the basis for the most advanced oscillator noise theories. Details can be found in \cite{Farkas}, while \cite{Demir4} deals with the extension of the classical Floquet theory of ODEs to DAEs.

Let us consider a LTV homogeneous ODE of size $n$
\begin{equation}
\totder{\yvec}{t}=\Amat(t)\yvec(t)
\label{simba1}
\end{equation}
where $\Amat(t)=\Amat(t+T)$ is a $T$-periodic matrix of size $n$. Given the initial condition $\yvec(0)=\yvec_0$, Floquet theorem states that the solution of \eqref{simba1} reads
\begin{equation}
\yvec(t)=\Smat(t,0)\yvec_0
\label{simba3}
\end{equation}
where $\Smat(t,s)$, the \textit{state transition matrix} of the LTV system, is expressed as
\begin{equation}
\Smat(t,s)=\Umat(t)\Dmat(t-s)\Vmat(s),
\end{equation}
where $\Umat$ and $\Vmat$ are two $T$-periodic invertible square matrices of size $n$ such that $\Umat(t)=\Vmat^{-1}(t)$, while matrix $\Dmat(t)$ is a diagonal matrix:
\begin{equation}
\Dmat(t)=\diag{\exp(\mu_1 t), \dots, \exp(\mu_n t)  }.
\label{simba2}
\end{equation}
The $n$ complex numbers\footnote{Since the LTV system is real, if a complex Floquet exponent exists, its complex conjugate should also be part of the Floquet exponents set.} $\mu_i$ are the Floquet exponents (FEs) of \eqref{simba1}, while $\lambda_i=\exp(\mu_i T)$ are the corresponding Floquet multipliers (FMs). According to the FM definition, for each FM $\lambda_i$ an infinite set of FEs exists, namely $\mu_i+\gei k 2\pi$ where $k$ is an integer number: this splitting of the FEs is important when the exponents are calculated by means of frequency domain techniques, such as Harmonic Balance \cite{IJCTA}.

Denoting with $\uvec_i(t)$ (resp. $\vvec_i^\up{T}(t)$) the $i$-th column (resp. row) of $\Umat(t)$ (resp. $\Vmat(t)$), the two sets $\{\uvec_i(t)\}$ and $\{\vvec_i(t)\}$ both span the entire $\mathbb{R}^n$, and form a bi-orthogonal basis (see \eqref{biob}).   Furthermore:
\begin{itemize}
\item $\uvec_i(t)\exp(\mu_i t)$ is a solution of \eqref{simba1} with initial condition $\uvec_i(0)$. For this reason, $\uvec_i(t)$ is the \textit{direct Floquet eigenvector} associated to the $\mu_i$ FE of \eqref{simba1};
\item $\vvec_i(t)\exp(-\mu_i t)$ is a solution of the adjoint system associated to \eqref{simba1}, i.e.
\begin{equation}
\totder{\zvec}{t}=-\Amat^\up{T}(t)\zvec(t),
\label{simba4}
\end{equation}
with initial condition $\vvec_i(0)$. Correspondingly, $\vvec_i(t)$ is the \textit{adjoint Floquet eigenvector} associated to  $\mu_i$.
\end{itemize}

Considering the limit cycle $\xvec_\up{S}(t)$ solution of \eqref{nosc}, and the LTV system defined by a linearization of \eqref{nosc}, i.e.
\begin{equation}
\Amat(t)=\left. \nabla_\xvec \fvec(x)\right|_{\xvec_\up{S}(t)},
\label{iringa}
\end{equation}
it follows
\begin{equation}
\totder{\phantom{t}}{t} \totder{\xvec_\up{S}}{t}=\totder{\phantom{t}}{t} \fvec(\xvec_\up{S}(t))=\Amat(t) \totder{\xvec_\up{S}}{t}.
\end{equation}
Therefore, for the LTV system associated to an autonomous circuit, $0$ is always a FE (or, equivalently, $+1$ is always a FM) associated to the direct Floquet eigenvector $\D\xvec_s/\D t$. With no loss of generality, we assume $\mu_1=0$ and $\uvec_1(t)=\D\xvec_s/\D t$.

Due to the exponential dependence on $\mu_i$ of the solution of \eqref{simba1}, an oscillator has an asymptotically stable orbit if and only if all the FEs $\mu_i$ ($i=2,\dots,n$) have negative real part, or equivalently all the FMs $\lambda_i$ ($i=2,\dots,n$) have magnitude lower than 1.

The computation of the FEs and eigenvectors (direct and adjoint) is a fundamental step for Floquet-based oscillator noise analysis. Specifically, the adjoint Floquet eigenvector $\vvec_1(t)$, associated to $\mu_1=0$, is the so-called \textit{perturbation projection vector} that plays the man role in the assessment of phase noise \cite{Demir1,Djurhuus,Suvak}. Due to their importance for oscillator noise and for the assessment of the stability of limit cycles \cite{Cappelluti2013}, the  Floquet quantities have been the object of research for several years. In most of the cases, the computation is perfomed in time domain \cite{GuckenheimerSJSC,Lust,Demirmu,milanesi08}. However, efficient algorithms for the frequemcy domain evaluation, based on the harmonic balance technique, are proposed in \cite{Demirmu,IJCTA,milanesi09,IET,AEU,TCAD}.

\bibliographystyle{spphys}       

\bibliography{invitedTomas}

%
%

\end{document}